# Convolutional Neural Network Array for Sign Language Recognition using Wearable IMUs

Karush Suri, Rinki Gupta
*Electronics and Communication Engineering Department*
*Amity University Uttar Pradesh,*
Noida, India
karushsuri@gmail.com rgupta3@amity.edu

*Abstract*—Advancements in gesture recognition algorithms have led to a significant growth in sign language translation. By making use of efficient intelligent models, signs can be recognized with precision. The proposed work presents a novel one-dimensional Convolutional Neural Network (CNN) array architecture for recognition of signs from the Indian sign language using signals recorded from a custom designed wearable IMU device. The IMU device makes use of tri-axial accelerometer and gyroscope. The signals recorded using the IMU device are segregated on the basis of their context, such as whether they correspond to signing for a general sentence or an interrogative sentence. The array comprises of two individual CNNs, one classifying the general sentences and the other classifying the interrogative sentence. Performances of individual CNNs in the array architecture are compared to that of a conventional CNN classifying the unsegregated dataset. Peak classification accuracies of 94.20% for general sentences and 95.00% for interrogative sentences achieved with the proposed CNN array in comparison to 93.50% for conventional CNN assert the suitability of the proposed approach.

## I. INTRODUCTION

Translation of sign language has gained significant importance as one of the primary branches of gesture recognition [1, 2]. Sign language is used by deaf and mute people to convey their messages in the form of structured sentences and questions. The syntax of sign language may be used to develop intelligent models that can accurately translate sign language to verbal languages so that a signer may interact with a person who does not understand signing. Wearable Inertial Measurement Units (IMUs) play an important role in acquiring data that may be used as input for the intelligent models to enable electronic translation of sign language. Unlike other sensors such as surface Electromyography (sEMG), IMUs are easy to handle, economical and require minimum invasiveness. IMUs assess motion of the sensor by virtue of change in position and orientation and hence, they have been used in hand gesture recognition..

Study of hand gestures has gained significant importance in the past decade. With advancements in sensing technology, healthcare development, gesture recognition systems for robotics and human machine interfaces, assistive technology including sign language translation systems have received a lot of attention from the research community. Translation of signs is carried out by various methods including one-, two- and three-dimensional signals. These help in providing a comprehensive idea about the motion of the limb. The limb, being a dynamic entity, undergoes change in position and orientation, which can be assessed by means of IMUs. IMUs include accelerometers and gyroscopes that measure the motion in form of acceleration and angular velocity, respectively [3]. However, IMUs are subject to directional components when the motion tends to become more complex. This requires an increase in the degrees of freedom or the number of sensors covering limb motion in Cartesian space [4]. Apart from limb tracking and motion assessment, other major application areas of IMU usage involve real-time scenarios in literature [5]. This is observed as a result of less computational expense and higher sampling rates in sensors. Other wearable sensors that have been used commonly in literature for hand motion analysis include sEMG [6], flex sensors [7], compared to which, based on our experience, IMU are most repeatable.

Once the data is recorded using wearable sensors, the next step is to develop intelligent models for recognition of the activity from the recorded signals. In recent years, deep models having layered architectures with pre-training capabilities are abundantly being used in literature [8]. Networks such as Artificial Neural Networks (ANNs) resembling the topology of neurotic systems in the human body are primarily adopted for recognition [8]. Increasing the depth of ANNs in terms of layers gives rise to deeper structures, commonly known as Deep Neural Networks (DNNs), capable of learning more intricate aspects of data [9]. Classification accuracies as high as 85-95% have been reported in literature when ANNs and DNNs are applied on IMU signals for activity recognition [10-11]. However, due to their layered architecture and the provision of back-propagation algorithm used for gradient optimization, ANNs and DNNs often undergo vanishing gradients which causes these models to stop learning. Furthermore, repetitive inputs to these networks also decreases their performance. This shortcoming of ANNs and DNNs is handled by Recursive Neural Networks (RNNs) [12]. RNNs, as a result of their feedback mechanism, are reported intelligent for sequence modeling. These are capable of recognizing sequential data and generating synthetic samples as well [13]. Although RNNs predominantly yield accurate predictions for signals and images, complex computation at the recursive layers along with the loss of data at long short-term memory (LSTM) cells makes RNNs less compatible for pre-training. Thus, recognition of input data requires an effective architecture capable of providing accurate results and convenient training techniques. One of the alternatives is the usage of Convolutional Neural Networks (CNNs) [14]. CNNs have been shown to be very useful in applications involving image processing for diagnostics development and human-machine cooperation in the process of disease detection. CNNs learn feature maps corresponding to the given input by making use of successive convolutional and pooling layers [15]. Presence of fully-connected layers act as

Funding support received from SERB, DST (ECR/2016/000637).

the recognition network, as previously observed in the case of ANNs and DNNs. Although CNNs are the most frequently used deep learning architecture in literature, their usage on 1D signals is limited and requires additional data for learning as the number of classes are increased [16]. Such a scenario increases the load on the network and diminishes its performance upon validation.

The work presented here proposes a novel one-dimensional (1D) CNN array consisting of two CNNs for the recognition of Indian Sign Language using data recorded by means of a custom designed wearable IMU system. The two CNNs forming the array operate in a sequential manner based on the input provided from the IMU dataset. The first CNN, hereby referred to as CNN-sentence, recognizes general sentences that are present in the considered dataset whereas the second CNN, CNN-questions, recognizes interrogative sentences from the dataset. Grouping of sentences and questions is carried out on the basis of a pre-defined structure in which signs indicating a subject or an object of the sentence are connected to each other. Performance of both the CNNs in the proposed CNN array is compared to a conventional CNN recognizing sentences and questions together, without making any distinction between them. The details of the recorded database and the proposed CNN array are contained in Section II. Results for evaluation of the performance of the proposed architecture are given in Section III. Section IV contains the concluding remarks.

## II. SENTENCE RECOGNITION USING CNN ARRAY

### A. Data Corpus

Fig. 1a shows the IMU GY-80 multi-board consisting of tri-axial accelerometer ADXL345 and gyroscope L3G400D used for measuring acceleration (in $m/s^2$) and turn rate (in deg/s) at 100 Hz. Both the instruments have 3 degrees-of-freedom each, giving a total of 6 degrees-of-freedom for the complete setup. Fig. 1b shown the complete experimental setup used in the recording of IMU signals using the GY-80 multiboard and Arduino UNO board having ATmega328P microcontroller. The setup is placed on wearable bands including straps. This is done for convenient usage of the apparatus and a compact design which would prevent any hindrances during hand motion. The setup is placed on the frontal side of the forearm, below the elbow.

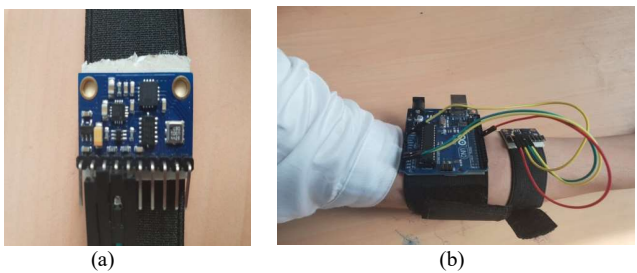

(a)      (b)

Fig. 1. IMU device for data recording (a) GY-80 sensor board, (b) Placement of experimental apparatus

IMU signals were recorded from a total of 10 different subjects, out of which 5 were male and 5 are female. All the subjects were in the age range of 21-49 years with 2 subjects being left-handed and the rest right-handed. Signs were recorded from the Indian Sign Language in a sequence, resulting in the formation of a complete sentence [17]. Each subject performed 20 sentences and 10 repetitions of each sentence. A uniform structure has been used for the sentences. Each sentence consists of 2-4 signs with a pause of around 1s between each sign. Hence, the sentence may be viewed as a sequence of connected signs. The signs are neither too far apart to be labeled as isolated signs nor signed fluidly as in continuous signing. This was done to reduce the complexity of recognition as compared to continuous signing. Interrogative words, number quantities and negations are placed as the last sign of the sentence whereas subjects and objects are placed as the first. For instance, for the sentence 'I need Help' gesture corresponding to 'I' is performed first followed by gesture corresponding to 'need' followed by gesture corresponding to 'Help'. However, in the sentence 'Where is the Doctor?' gesture corresponding to 'Doctor' is performed first followed by gesture corresponding to 'Where'. Thus, there exists a clear distinction between signals of sentences and questions corresponding to each subject.

### B. Array Architecture

The proposed array architecture as depicted in Fig 2. is a 2x1 array consisting of 2 CNNs. Each CNN operates on a particular subset of data, i.e. either questions or sentences. These subsets have been manually created prior to the training phase and split into training and validation data. Both the CNNs have identical structure for convenience. Each network consists of 9 layers, these being convolutional layers, pooling layers, flattening layers and fully-connected perceptron layers. The convolution layer consists of 128 and 256 filters respectively, each having a size of 5x5 and activation as Regularized Tangent (Relu) The pooling subsample space has a pool size of 3. Convolutional and pooling layers together make the feature learning network. The recognition network consists of fully-connected (FC) multi-perceptron layers. The first layer in the recognition network consists of 128 hidden units whereas the number of units in the final layer depend on the number of classes. Since there are 12 general sentences in the IMU dataset, CNN-sentence consists of 12 units in the final layer. Similarly, 8 questions in the IMU dataset make up for 8 units in CNN-questions. Each of the output layers in the CNN array are activated by means of sigmoid activation. Flattening of learned entities is carried out for the purpose of dimensionality reduction followed by predictions offered by the recognition network. Optimizer used for the process of recognition is *RMSprop* initialized to a learning rate of 0.0001 and no decay. Operation of the array is governed by sequential classification of input data. IMU dataset is subdivided into subsets of sentences and questions. Each of these subsets consists of data from all the 10 subjects having 10 repetitions of each class. Thus, the sentence subset has a total of 10*10*12 = 1200 signals as samples (corresponding to 12 classes) whereas the questions subset has 10*10*8 = 800 signals as samples (corresponding to 8 classes). Each of these subsets is further split into training and validation data with 90% of the samples falling in the training set. CNN-sentences is therefore trained on 1080 signals and validated on 120 signals whereas CNN-questions is trained on 720 signals and validated on 80 signals. Since there are multiple classes for each CNN in the array, the use of categorical cross entropy as a suitable loss function is employed.

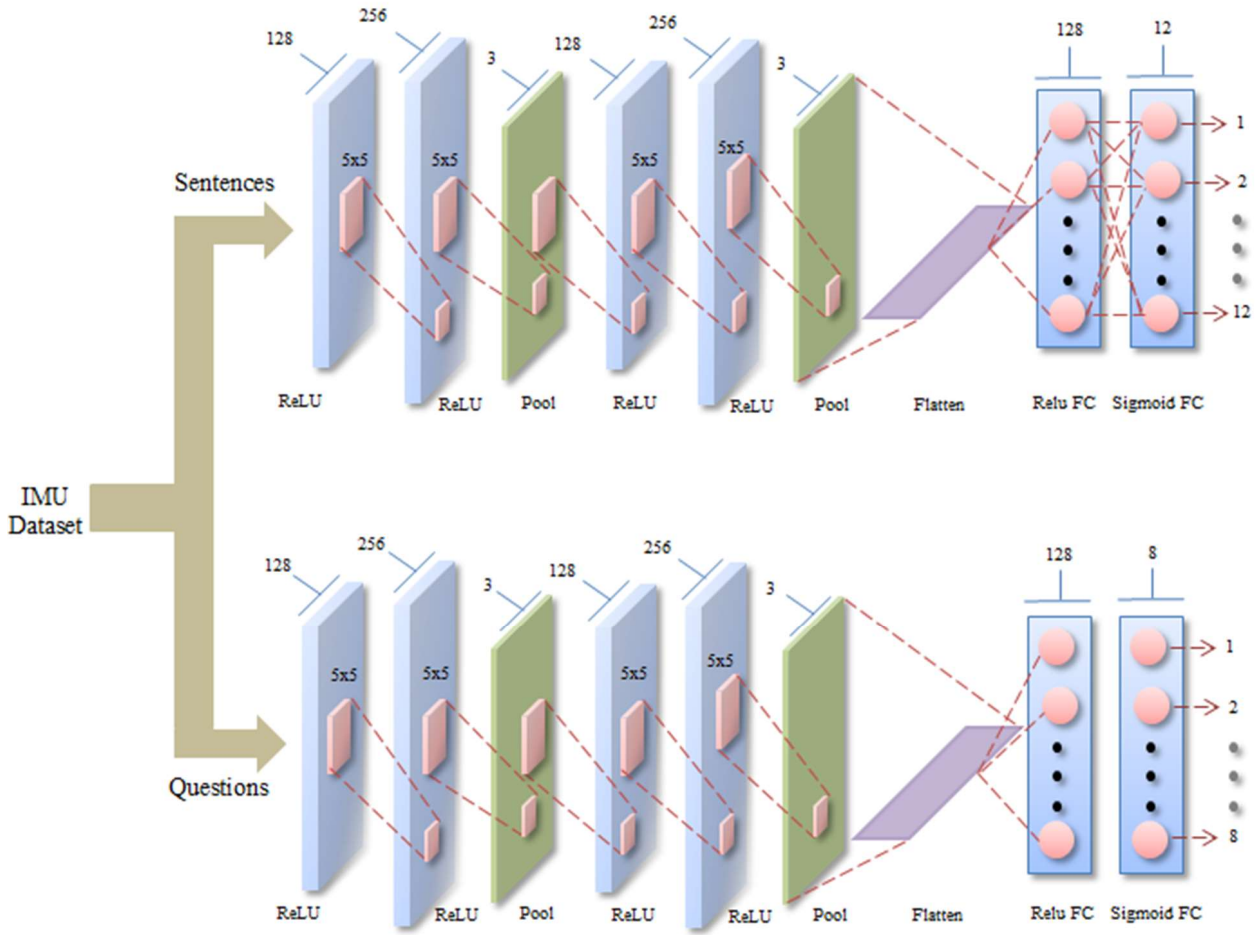

Fig. 2. Proposed CNN Array Architecture for Sign Language Recognition

Equations (1) and (2) denote the expression for categorical cross entropy on the total number of samples 'n' in a training dataset.

$$H(p,q) = -[(p(x_1)\log q(x_1)) + (p(x_2)\log q(x_2)) + \ldots (p(x_n)\log q(x_n))], \quad (1)$$

$$H(p,q) = -\sum_{i=1}^{n}[(p(x_i)\log q(x_i))]. \quad (2)$$

Here, $x_i$ denotes the $i^{th}$ sample in the subset, $p(x_i)$ the class probability corresponding to the $i^{th}$ sample and $q(x_i)$ the likelihood corresponding to the $i^{th}$ sample. The loss in multi-layered networks may not depend on incorrect classes. However, the gradient of the loss function does affect incorrect classes. This leads to an optimization in the learned weights over each passing iteration. Gradient of the loss function can be expressed as the derivative with respect to the $i^{th}$ sample as given in (3),

$$\frac{\partial H(p,q)}{\partial x_i} = -\frac{\partial[p(x_i)\log(q(x_i))]}{\partial x_i}. \quad (3)$$

This can further be simplified to give,

$$\frac{\partial H(p,q)}{\partial x_i} = -p(x_i)\frac{1}{q(x_i)}\frac{\partial q(x_i)}{\partial x_i} \quad (4)$$

Equation (4) indicates the gradient updates taking place for each sample in the training set. True classes representing a finite value of $p(x_i)$ are taken into account. Values corresponding to false classes represent a zero value of $p(x_i)$. Categorical representation of predictions is thus, organized as 1's and 0's indicating true and false classes, respectively.

### III. RESULTS AND DISCUSSION

The proposed CNN architecture is applied on the IMU dataset to classify the signed sentences, as explained in Section II. For comparison, the conventional CNN with no segregation distinction between general sentences and questions is also applied on the same dataset and the results for classification are compared with the proposed approach. The proposed 2x1 1D array architecture results in an improvement of the recognition process when compared to the conventional CNN. Figures 3a and 3b depict the variation of average classification accuracy over 50 iteration for training and validation phases, respectively. The classification accuracies improve as the number of epochs increase and finally stabilizes at around 40 epochs. As seen in Fig. 3, at the end of 50 epochs, the classification accuracies achieved by all the three networks is above 90%. However, both the networks in the array, CNN-sentences and CNN-questions depict an increase in the performance, given same hyper-parameters and constraints for validation to both the models. The CNN-questions model achieves a higher classification accuracy in comparison to the accuracies achieved using CNN-sentences and CNN-conventional.

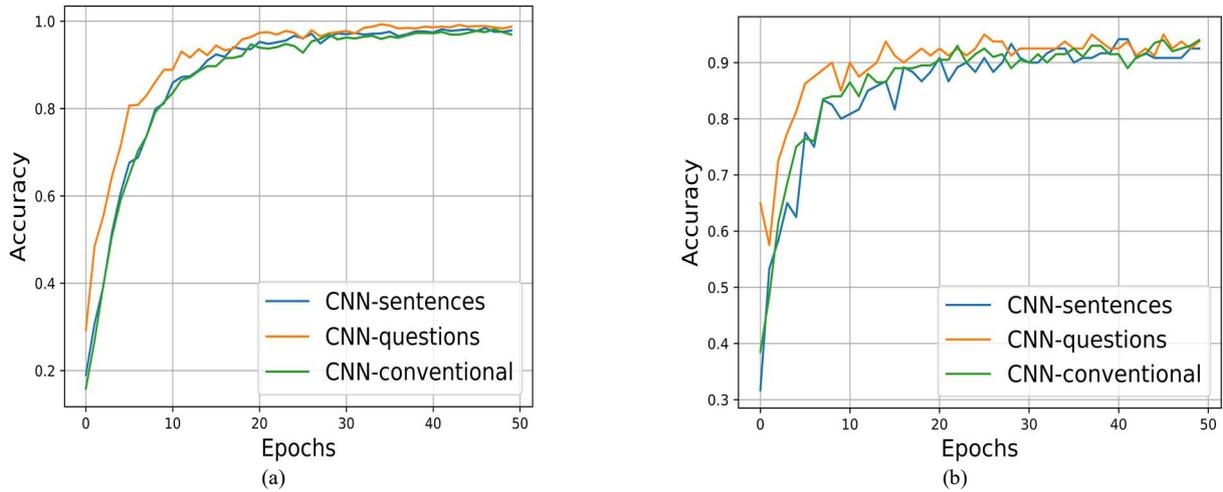

Fig. 3. Variation of classification accuracy measured over 50 iterations (a) Training Accuracy, (b) Validation Accuracy

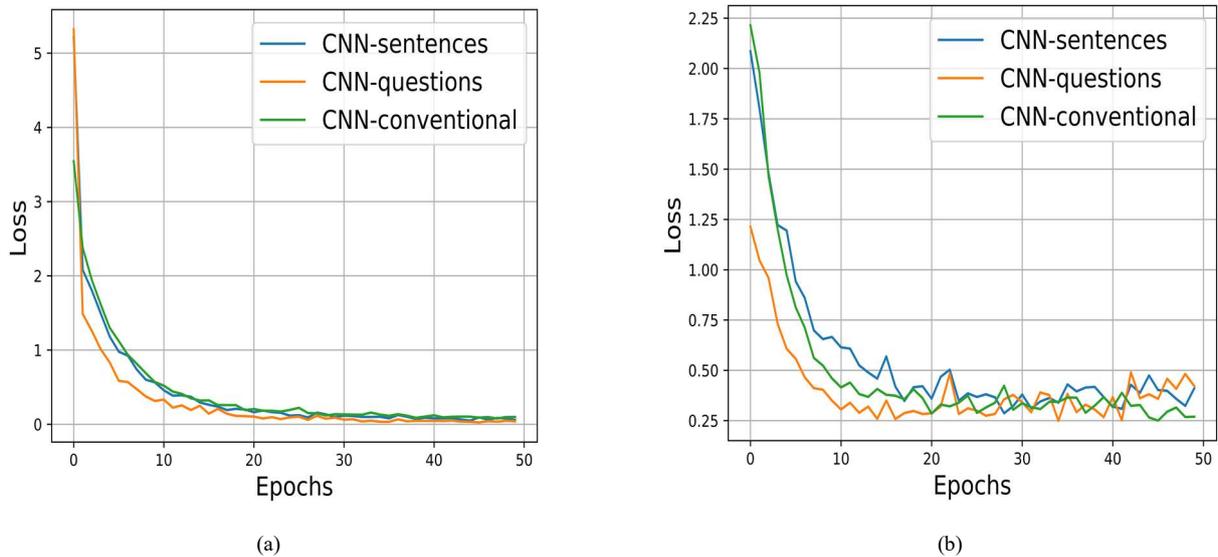

Fig. 4. Variation of Loss measured over 50 iterations (a) Training Loss, (b) Validation Loss

The variation of optimization losses of the considered models is compared during training and validation phases in Figs. 4a and 4b, respectively. The minimal gradient optimization is observed in the case of CNN-questions is observed in comparison to the optimization for CNN-sentences and CNN-conventional. Figures 5a and 5b represent the variation of false predictions offered by models during training and validation phases respectively. False predictions are employed as a validation metric here primarily for the assessment of optimization loss. A lower gradient initialization at the start of each epoch may not indicate fewer misclassified entities. Thus, falsely predicted values are plotted separately for the evaluation of dynamic optimization. As seen in Fig. 5a and 5b, false predictions for CNN-questions depict minimum misclassified entities followed by CNN-sentences and CNN-conventional. Thus, corresponding to each metric used for training and validation, CNN-questions presents peak variation of improved recognition followed by CNN-sentences and CNN-conventional. Improvement of CNNs in the 2x1 array in comparison to CNN-conventional assert the suitability of the architecture.

Table I highlights the peak performance values for all the metrics used to assess the models at the end of 50 iterations. Optimization loss presents a peak value of 0.02 and 0.24 during training and validation in the case of CNN-questions which is the highest among the three models followed by CNN-sentences and CNN-conventional. The optimization loss for validation phase are comparable. This may be due to the limited amount of data used for validation. In future, the database will be expanded to include more subjects and more sentences.

As seen in Table I, the number of false predictions is fewer in the case of CNN-questions (5 for training and 4 for validation) in comparison to CNN-sentences (17 for training and 7 for validation) and CNN-conventional (33 for training and 12 for validation). Average classification accuracy values depict accurate recognition values of 99.30% and 95.00% for CNN-questions during training and validation, followed by 98.42% and 94.20% for CNN-sentence during training and validation and 98.16% and 93.50% for CNN-conventional during training and validation. Higher peak performance values in the case of CNN-questions and CNN-sentences in comparison to CNN-conventional depict the suitability of the proposes array architecture approach.

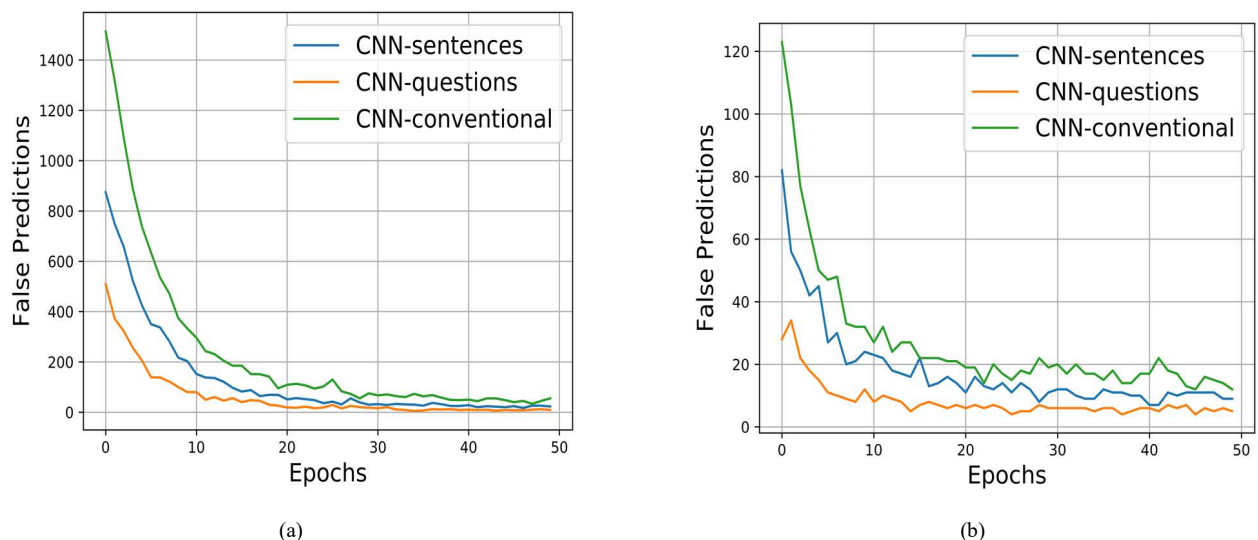

Fig. 5. Variation of False Prediction measured over 50 iterations (a) During Training, (b) During Validation

TABLE I. PEAK PERFORMANCE VALUES FOR IMU SIGNAL RECOGNITION (OBTAINED OVER 50 ITERATIONS)

| Model | Optimization Loss | | False Predictions | | Classification Accuracy | |
|---|---|---|---|---|---|---|
| | Training | Validation | Training | Validation | Training | Validation |
| CNN-sentences | 0.04 | 0.24 | 17 | 7 | 98.42% | 94.20% |
| CNN- questions | 0.02 | 0.24 | 5 | 4 | 99.30% | 95.00% |
| CNN- conventional | 0.07 | 0.25 | 33 | 12 | 98.16% | 93.50% |

## IV. CONCLUSION

Recognition of sign language has gained popularity as one of the primary applications of gesture recognition. Intelligent algorithms in corroboration with wearable sensors pave the way for accurate recognition of signs in the form of sentences. The work proposed here presents a novel one-dimensional CNN array architecture for the recognition of sentences signed according to the Indian Sign Language. Signals for the signed sentences are recorded by using a custom designed wearable IMU device. The dataset is split into general sentences and questions. Two CNNs in the array recognize the general sentences and questions separately. Performance of the proposed array architecture is compared to a conventional CNN which is trained to recognize any sentence from the complete dataset. The number of false predictions is less for CNN-array, which is 17 and 5 for CNN-sentences and questions, respectively, as compared to conventional CNN where it is 33 during training. During validation, peak classification accuracy values of 95.00% for CNN-questions and 94.20% for CNN-sentences in the deep array architecture in comparison to 93.50% for the conventional CNN validate the suitability of the proposed approach. In future we will expand the database to include more subjects and more sentences, as well as decrease the pause between signs to achieve continuous sign language recognition.

ACKNOWLEDGMENT

The author would like to thank the volunteers who helped in recording the data. The author also recognizes the funding support provided by the Science & Engineering Research Board, a statutory body of the Department of Science & Technology (DST), Government of India (ECR/2016/000637).